# Beyond Parrondo's Paradox

**Jian-Jun SHU* and Qi-Wen WANG**

School of Mechanical and Aerospace Engineering, Nanyang Technological University,

50 Nanyang Avenue, Singapore 639798

**The Parrondo's paradox is a counterintuitive phenomenon where individually-losing strategies can be combined in producing a winning expectation. In this paper, the issues surrounding the Parrondo's paradox are investigated. The focus is lying on testifying whether the same paradoxical effect can be reproduced by using a simple capital dependent game. The paradoxical effect generated by the Parrondo's paradox can be explained by placing all the parameters in one probability space. Based on this framework, it is able to generate other possible paradoxical effects by manipulating the parameters in the probability space.**

The Parrondo's paradox describes the counterintuitive situation where combining two individually-losing games could produce a winning expectation. The initial purpose of the Parrondo's paradox was to simulate a counterintuitive physical phenomenon generated by the flashing Brownian ratchet[1] in terms of two gambling games[2]. Some studies were made to demonstrate the concept of the capital-dependent Parrondo's paradox[3,4], to formulate the mathematical expressions of the Parrondo's paradox[5,6], and to extend the capital-dependent Parrondo's paradox to a history-dependent version[7].

The Parrondo's paradox has raised attention as it has tremendous potentials in describing the strategy of turning two unfavorable situations into a favorable one. The concept has been scrutinized[8,9] since its first appearance and extended into other potential applications[10-13].

In this paper, it begins with a short summarization of key concepts of the Parrondo's paradox. It is further ventured into the investigation on whether the analogous paradoxical effect can be reproduced by using a relatively simple capital-dependent game as claimed before[14]. In reality, all the parameters that used in the Parrondo's paradox can be analyzed in a probability space, which reveals the working principle of the paradox. Based upon this foundation, it is possible to generate other paradoxical effects by manipulating these parameters inside the probability space. In the end, the issues associated with paradox are discussed.

---

* Correspondence should be addressed to Jian-Jun SHU, mjjshu@ntu.edu.sg

There are totally two versions of the Parrondo's paradox, which is referred to as capital- and history-dependent. The Parrondo's paradox consists of two games, namely, game A and game B. The only difference between these two versions of paradox is lying on the corresponding switching mechanisms of game B. For both versions, game A is exactly the same. It is a zero-order memoryless gambling game of winning probability of $p_1$ and losing probability of $1-p_1$. Game B is a condition-based game, also known as the second-order Markov game, which consists of two scenarios – *scenario 1* and *scenario 2*.

For the capital-dependent Parrondo's paradox, choosing which scenario to be played merely depends upon whether the instantaneous capital $C(t)$ is a multiple of predefined integer $M$ or not. If the capital $C(t)$ is a multiple of $M$ (*i.e.* $C(t) \bmod M = 0$), *scenario 1* is chosen to be played, in which the winning probability $p_2$ is much lower than the losing probability $1-p_2$. If the capital $C(t)$ is not a multiple of $M$ (*i.e.* $C(t) \bmod M \neq 0$), *scenario 2* is selected, in which the winning probability $p_3$ is slightly higher than the losing probability $1-p_3$.

For the history-dependent Parrondo's paradox, deciding which scenario to be played relies on the outcomes of previous two games. As the outcome of each game is resulting in a win or loss, there are totally four different combinations of results of previous two games: {lose, lose}, {lose, win}, {win, lose} and {win, win}. Therefore, there are totally four different scenarios to be selected. Each scenario corresponds to one specific combination of results of previous two games.

Three probabilities, $p_1$, $p_2$ and $p_3$, are controlled by using one single biasing parameter $\varepsilon$. The central idea is that, by setting biasing parameter $\varepsilon > 0$, both game A and game B are losing games (*i.e.* capital $C(t)$ is decreasing with the advancement of number $t$ of games played) if played individually. The Parrondo's games are illustrated in Figure 1.

## Capital-dependent Parrondo's paradox

### Game A

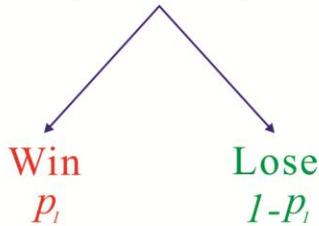

| Game A | |
|---|---|
| $p_1$ (Win) | $1-p_1$ (Lose) |
| $0.5 - \varepsilon$ | $0.5 + \varepsilon$ |

### Game B

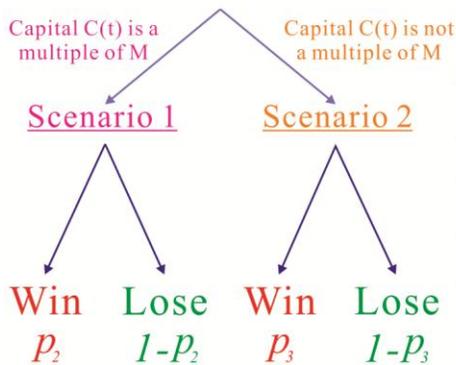

| Game B | | | |
|---|---|---|---|
| Is capital $C(t)$ a multiple of M? | | | |
| Scenario 1 (Yes) | | Scenario 2 (No) | |
| $p_2$ (Win) | $1-p_2$ (Lose) | $p_3$ (Win) | $1-p_3$ (Lose) |
| $0.1 - \varepsilon$ | $0.9 + \varepsilon$ | $0.75 - \varepsilon$ | $0.25 + \varepsilon$ |

## History-dependent Parrondo's paradox

### Game A

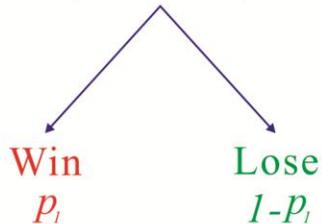

| Game A | |
|---|---|
| $p_1$ (Win) | $1-p_1$ (Lose) |
| $0.5 - \varepsilon$ | $0.5 + \varepsilon$ |

### Game B

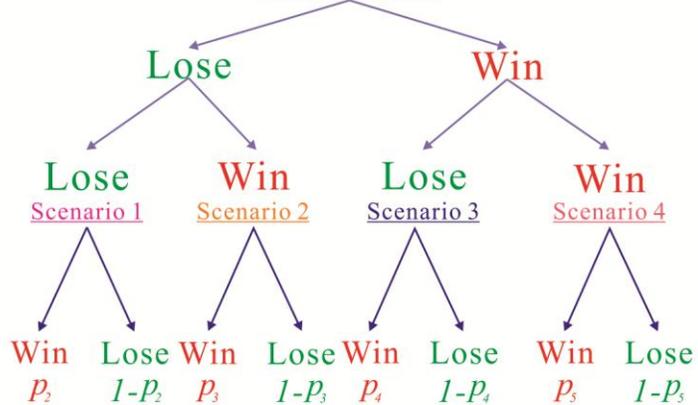

| Game B | | | | | | | |
|---|---|---|---|---|---|---|---|
| Lose | | | | Win | | | |
| Lose Scenario 1 | | Win Scenario 2 | | Lose Scenario 3 | | Win Scenario 4 | |
| $p_2$ (Win) | $1-p_2$ (Lose) | $p_3$ (Win) | $1-p_3$ (Lose) | $p_4$ (Win) | $1-p_4$ (Lose) | $p_5$ (Win) | $1-p_5$ (Lose) |
| $0.9 - \varepsilon$ | $0.9 + \varepsilon$ | $0.25 - \varepsilon$ | $0.25 + \varepsilon$ | $0.25 - \varepsilon$ | $0.25 + \varepsilon$ | $0.7 - \varepsilon$ | $0.7 + \varepsilon$ |

**Figure 1**: *Parrondo's games*

Based on the rules of games as specified in Figure 1 and by setting the value of biasing parameter $\varepsilon = 0.005$ and predefined integer $M = 3$, respectively, a simulation can be generated by averaging the outcomes of *10,000* trials for each game, and totally *200* games are played, as shown in Figure 2.

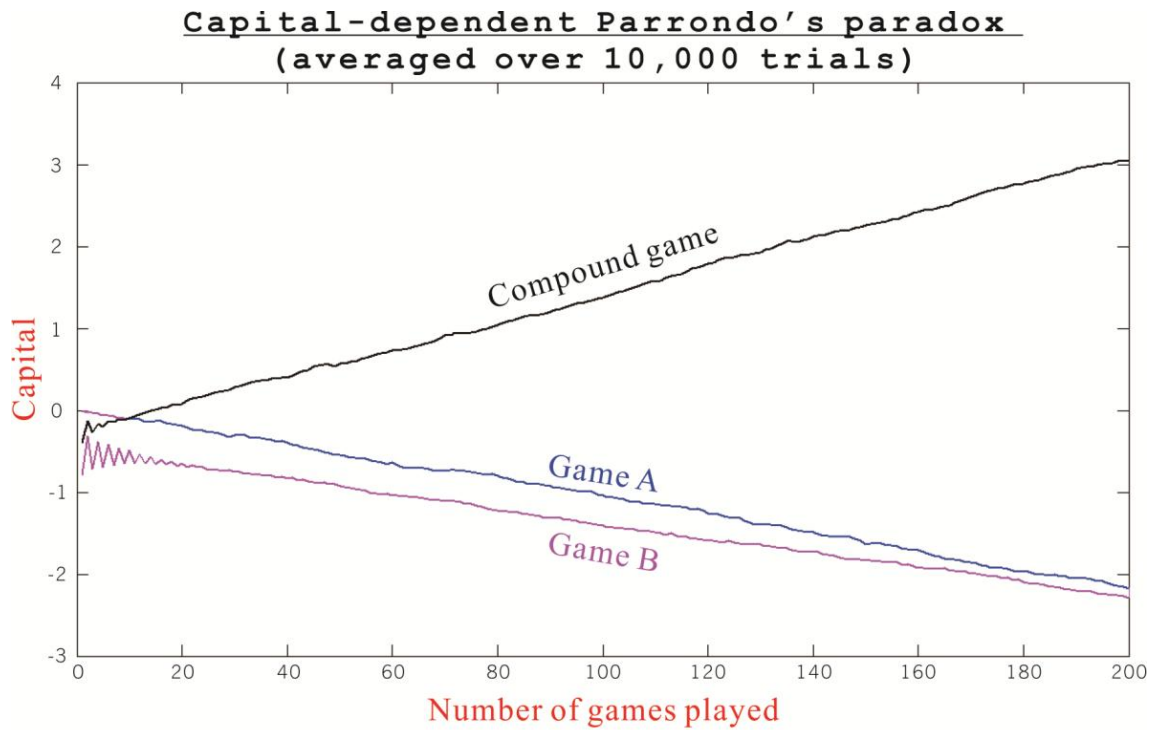

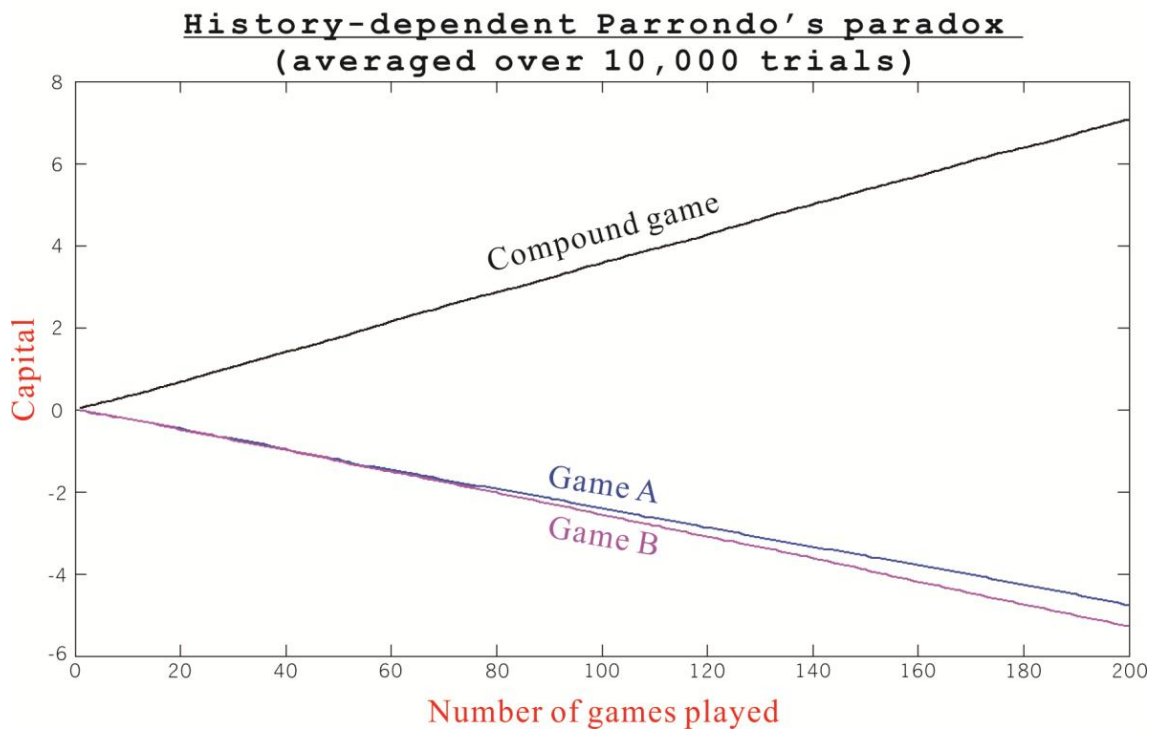

**Figure 2**: *Simulation of Parrondo's paradox*

Figure 2 reveals two essential information: for the capital-dependent Parrondo's paradox, both game A (blue) and game B (pink) are losing games if played individually; however, once these two games are played in a mixed manner, in which both game A and game B have equal chance to be played (*i.e. Probability*(*game* A) = *Probability*(*game* B)), the resultant compound game (black) is a winning game.

## Results

The counterintuitive phenomenon, which is generated by the compound game, or randomly mixed game, of the capital-dependent Parrondo's paradox, can be analyzed by simply placing all the probabilities in one single probability space[15]. Such a probability space, as shown in Figure 3, consists of two elements: a straight line (red) and a curve (black).

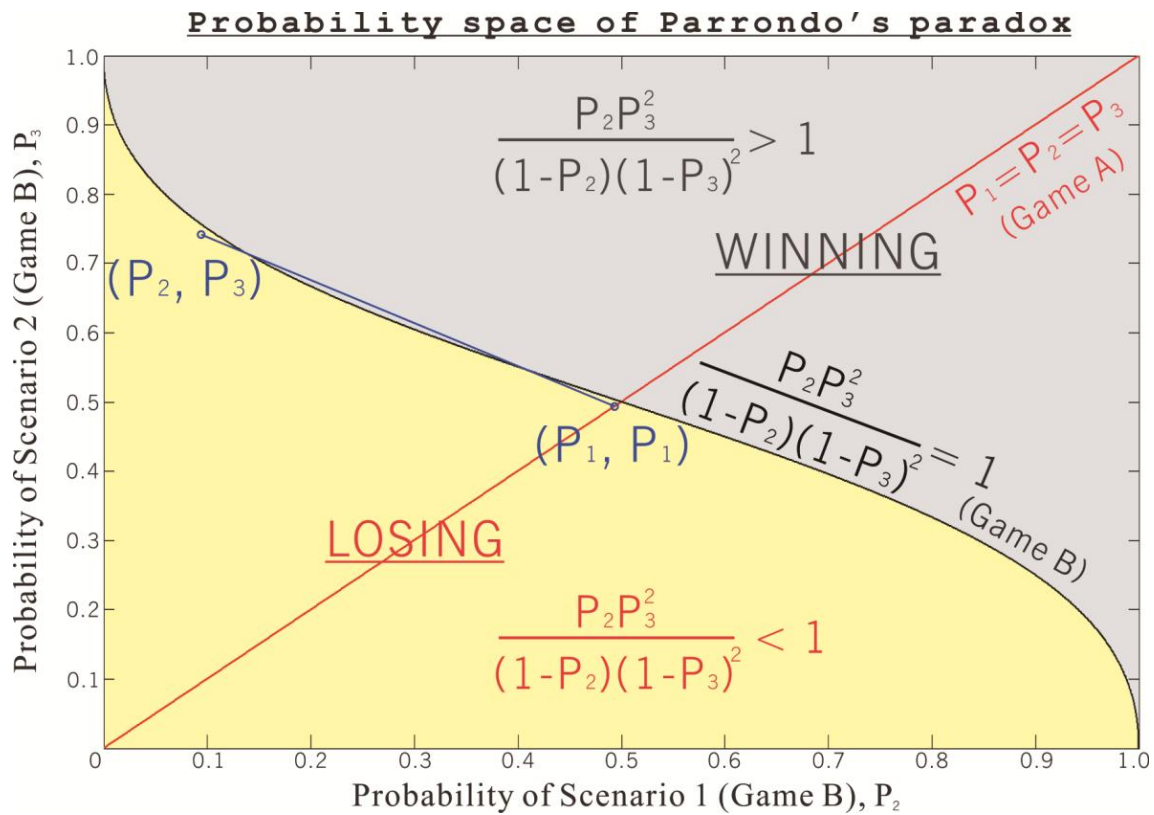

**Figure 3**: *Probability space of capital-dependent Parrondo's paradox*

The curve is specified by the game rules of game B. In order to make game B a fair game, the winning probability must equal to the losing probability, that is, $p_2 p_3^{M-1} = (1-p_2)(1-p_3)^{M-1}$. In the selected case of the capital-dependent Parrondo's paradox, predefined integer is selected to be $M = 3$. Therefore, in order to make the game B of the capital-dependent Parrondo's paradox a fair game, the probabilities of *scenario 1* and *scenario 2* of game B must satisfy equation (1) as indicated below.

$$\frac{p_2 p_3^2}{(1-p_2)(1-p_3)^2} = 1. \tag{1}$$

In addition, equation (1) can be modified by simply expressing the winning probability of *scenario 1*, $p_2$, in terms of the winning probability of *scenario 2*, $p_3$. The resultant function is equation (2), which is the curve (black) in Figure 3.

$$p_2 = \frac{(1-p_3)^2}{(1-p_3)^2 + p_3^2}. \tag{2}$$

It divides the entire probability space into two separate regions: the region above the curve is termed as *winning region* (grey) due to the winning probability of game B is higher than the corresponding losing probability, *i.e.*, $p_2 p_3^2 > (1-p_2)(1-p_3)^2$; the region below the curve is termed as *losing region* (yellow) due to the winning probability of game B is lower than the corresponding losing probability, *i.e.*, $p_2 p_3^2 < (1-p_2)(1-p_3)^2$.

In short, if the selected probabilities of game B, $p_2$ and $p_3$, are falling into the *winning region*, game B is a winning game. On the other hand, if the selected probabilities of game B, $p_2$ and $p_3$, are lying inside the *losing region*, game B is a losing game.

Similarly, by setting probabilities $p_1 = p_2 = p_3$, equation (1) can be converted into equation (3) as stated below.

$$p_1^3 = (1-p_1)^3. \tag{3}$$

By solving equation (3), it returns with three solutions: one real solution $p_1 = \frac{1}{2}$, and two imaginary solutions $p_1 = \frac{1}{2} - \frac{\sqrt{3}}{2}i$ and $p_1 = \frac{1}{2} + \frac{\sqrt{3}}{2}i$. The real solution implies the winning probabilities equals to the losing probability of game A. Such a relationship can be expressed in terms of a straight line (red) in the probability space, as shown in Figure 3. The winning probability of game A, $p_1$, is selected along this straight line. If the winning probability of game A is $p_1 < \frac{1}{2}$, the part of the straight line falls in the *losing region* and then game A is a losing game. On the other hand, if the winning probability of game A is larger than $p_1 > \frac{1}{2}$, the part of straight line falls into the *winning region* and then game A is a winning game. If the winning probability of game A is $p_1 = \frac{1}{2}$, the intersection point of the straight line and the curve and then game A is a fair game.

As specified in the game rules of the original capital-dependent Parrondo's paradox, game A is a losing game. Therefore, the winning probability $p_1$ of game A is selected along the straight line (red) in the *losing region*. In the original capital-dependent Parrondo's paradox, the winning probability of game A is $p_1 = 0.495$. Game B is also a losing game and therefore two winning probabilities of game B, $p_2$ and $p_3$, must be any single point, ($p_2$, $p_3$), located inside the *losing region*. The selected probabilities of the original Parrondo's paradox are $p_2 = 0.095$ and $p_3 = 0.745$. Therefore, it is possible to plot these two points, ($p_1$, $p_1$) = (0.495, 0.495) and ($p_2$, $p_3$) = (0.095, 0.745), inside the probability space.

The compound game is formed as a convex linear combination of two games, game A and game B, by introducing one additional parameter, namely, mixing parameter, denoted by $\gamma$. The parameter $\gamma$ is defined as the probability of selecting game A. Analogous to the game B of the capital-dependent Parrondo's paradox, the compound game is also a condition-based game. Suppose the capital $C(t)$ is divisible by $M$, the winning probability $p_{c1}$ of compound game can be expressed by equation (4).

$$p_{c1} = \gamma \, p_1 + (1-\gamma) p_2. \tag{4}$$

On the other hand, if the capital $C(t)$ is not a multiple of $M$, the winning probability $p_{c2}$ of compound game can be expressed by equation (5).

$$p_{c2} = \gamma \, p_1 + (1-\gamma) p_3. \tag{5}$$

As all these probabilities are fixed once they are selected, the only method is to tweak the value of mixing parameter $\gamma$. In order to make the compound game a winning one, the selected mixing parameter must satisfy equation (6), that is, the winning probabilities of compound game is greater than the corresponding losing probabilities if the predefined integer $M = 3$.

$$\frac{p_{c1} p_{c2}^2}{(1-p_{c1})(1-p_{c2})^2} > 1. \tag{6}$$

Such a method can also be represented in the same probability space, as shown in Figure 3, by linking these two points, ($p_1$, $p_1$) and ($p_2$, $p_3$), using a straight connecting line (blue). It is able to observe there is a certain region of the line falling inside the *winning region*. The probabilities fall inside this region satisfy equation (6), which makes compound game a winning one. By adjusting the value of mixing parameter $\gamma$, *i.e.*, changing the location of the point along the straight line, any selected points along this straight line falling into the *winning region* are the keys in producing a winning expectation. In the original capital-dependent Parrondo's paradox, mixing parameter $\gamma$ equals to $\frac{1}{2}$, which is the middle point of the straight line (blue). Such a point is located inside the *winning region*. Therefore, the compound game is a winning one.

Based upon the theoretical foundation, it is possible to construct several alternative designs, which can be used to explain how analogous paradoxical effect can be reproduced by simply manipulating parameters in the probability space.

It is started by providing a relatively simple alternative design, namely, the reversed Parrondo's paradox, that is, two individual winning games can also be combined in producing a losing expectation.

The reversed Parrondo's paradox is achieved by simply switching the winning probabilities with its losing probabilities. Similarly, the selected probabilities can be plotted inside the same probability space, as shown in Figure 4.

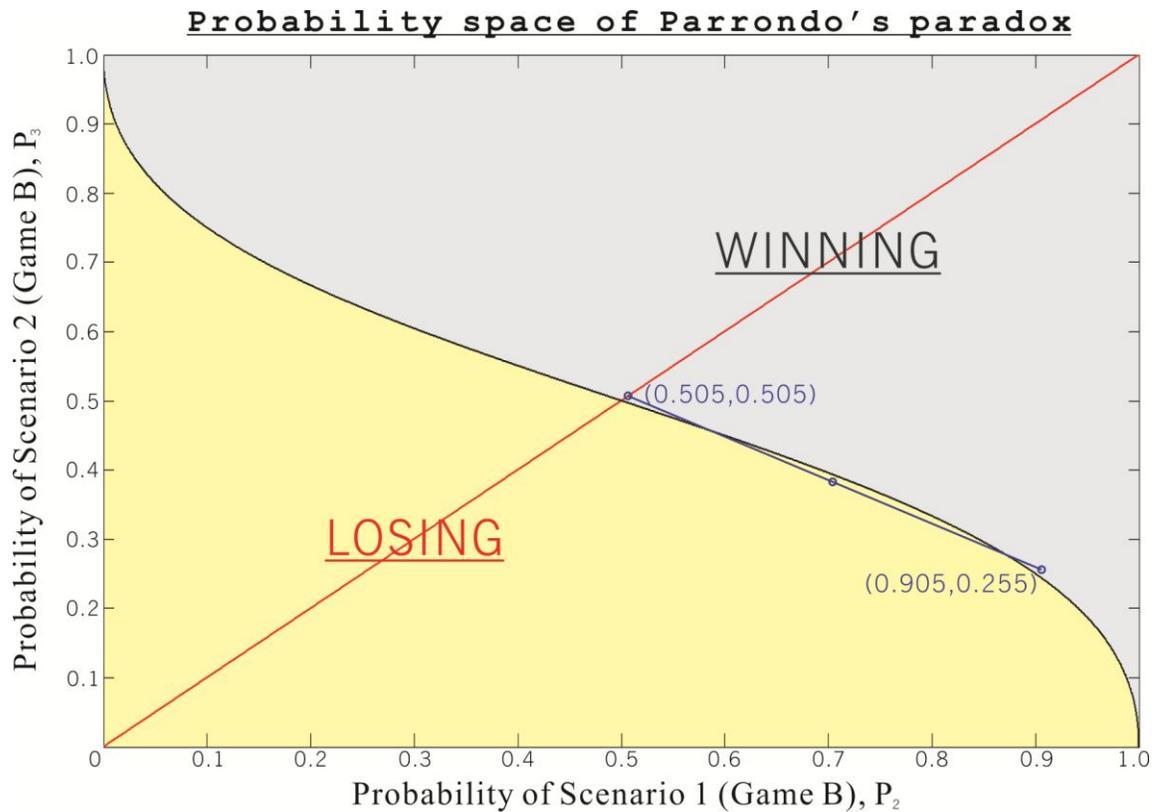

**Figure 4**: *Probability space of reversed Parrondo's paradox*

In the case, both probabilities of game A and game B are falling into the *winning region*. By setting the mixing parameter $\gamma$ to $\frac{1}{2}$, the middle point (of connecting line) is falling into the *losing region*. In the simple arrangement, it is to produce a totally reversed paradoxical effect.

The Parrondo's paradox is a combination of two losing games in producing one winning expectation. However, there are totally eight different combinations of two winning and/or losing games, including the Parrondo's paradox, which are summarized in Table 1.

**Table 1:** *Different combinations of two winning and/or losing games*

| Scheme | Game A | Game B | | |
|---|---|---|---|---|
| #1 | Lose | Lose | Scenario 1 | Lose |
| | | | Scenario 2 | Win |
| #2 | Win | Lose | Scenario 1 | Lose |
| | | | Scenario 2 | Win |
| #3 | Lose | Win | Scenario 1 | Win |
| | | | Scenario 2 | Win |
| #4 | Lose | Lose | Scenario 1 | Lose |
| | | | Scenario 2 | Lose |
| #5 | Win | Win | Scenario 1 | Win |
| | | | Scenario 2 | Win |
| #6 | Win | Lose | Scenario 1 | Lose |
| | | | Scenario 2 | Lose |
| #7 | Lose | Win | Scenario 1 | Win |
| | | | Scenario 2 | Lose |
| #8 | Win | Win | Scenario 1 | Win |
| | | | Scenario 2 | Lose |

*Scheme #1* is the Parrondo's paradox. Here the aim is to investigate whether the remaining seven combinations, from *scheme #2* to *#8*, are capable of producing other possible paradoxical effects. In order to preserve the consistence, the same value of biasing parameter $\varepsilon = 0.005$ and predefined integer $M = 3$ is used for all simulations. Analogous to the original version (*scheme #1*), a series of simulations in Figure 5 are generated by averaging the outcomes of *10,000* trials for each game, and totally *200* games are played.

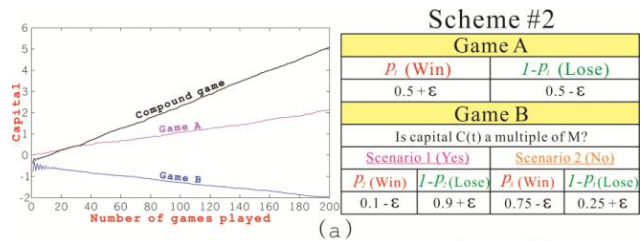
(a)

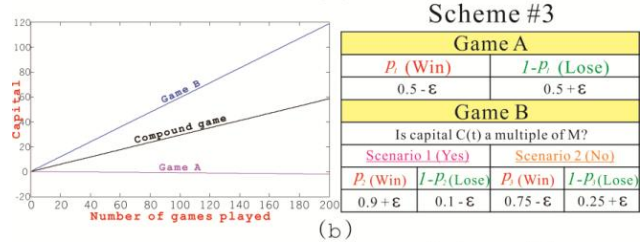
(b)

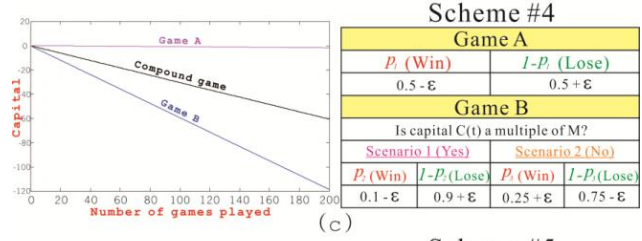
(c)

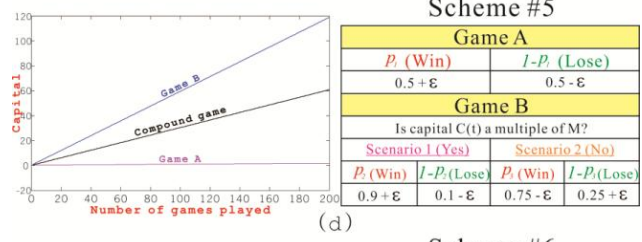
(d)

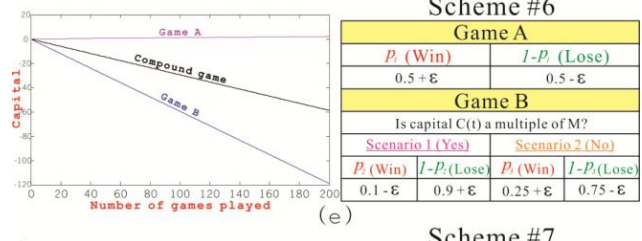
(e)

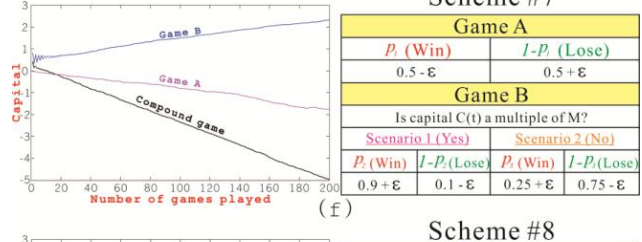
(f)

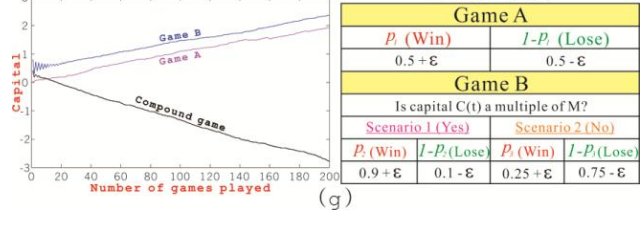
(g)

**Figure 5**: *Game rule and simulation results*

Both schemes *#4* (Figure 5(c)) and *#5* (Figure 5(d)) are belonging to trivial cases. In *scheme #4* (Figure 5(c)), the winning probabilities, $p_1$, $p_2$ and $p_3$, are smaller than the corresponding losing probabilities, $1-p_1$, $1-p_2$ and $1-p_3$. Therefore, there is no doubt that both game A and game B are losing games, and hence, the compound game is also a losing game. The same situation occurs in *scheme #5* (Figure 5(d)), both game A and game B are winning games. It is intuitive to have the compound game a winning game. In short, these two schemes, *#4* and *#5*, are not producing any paradoxical effects.

*Schemes #3* (Figure 5(b)) and *#6* (Figure 5(e)) are also regarded as trivial cases. In both schemes, game B is a complete either winning (*scheme #3*) or losing (*scheme #6*) game in both scenarios – *scenario 1* and *scenario 2*. The trend of compound game in each case is significantly influenced by that of game B in both schemes. In both schemes, *#3* (Figure 5(b)) and *#6* (Figure 5(e)), the instantaneous capital $C(t)$ at any number of games played is equal to half the sum of game A and game B. The generated phenomenon in both schemes is intuitive and, hence, they are not regarded as paradoxes.

On the other hand, schemes *#2* (Figure 5(a)) and *#7* (Figure 5(f)) produce relatively strong paradoxical effect. In *scheme #2*, game A is slightly winning game, game B is a complete losing game. Intuitively, the compound game should be a slightly losing game. However, as shown in Figure 5(a), the compound game is definitely outperformed game A, which results in a winning game. The identical situation also occurs in *scheme #7*, the only difference is that playing the compound game results in an obvious inferior position than game A alone.

Finally, the *scheme #8* (Figure 5(g)) is a complete reverse Parrondo's paradox, which produces a very strong paradoxical effect. In the original Parrondo's paradox (*scheme #1*), both game A and game B are losing games if played individually. The compound game of game A and game B, however, produces a complete counterintuitive phenomenon, resulting in a winning game. Similarly, in *scheme #8*, game A and game B are winning games if played individually. The compound game, as shown in Figure 5(g), is capable of producing a losing expectation.

## Discussion

From the scrutiny of the Parrondo's paradox[8,9], there are several issues surround the Parrondo's paradox since its first appearance. Some of these issues were responded by its initiators[16]. The objective of this paper is to resolve the remaining issues associated with the Parrondo's paradox. It begins by focusing on testifying whether the identical paradoxical effect can be simply reproduced by replacing a relatively simple capital-dependent game as claimed before[14], which also involves two games – game A: player loses *$2* if his capital $C(t)$ is an odd number, and loses *$1* if $C(t)$ is an even number; game B: the player gains *$6* if $C(t)$ is an odd number, and loses *$7* is $C(t)$ is an even number.

At first glance, the proposed game seems to be plausible. In order to verify whether the paradoxical effect could be generated by the proposed simple capital-dependent game, a simulation is presented in Figure 6. As indicated in Figure 6, game A is a losing game and the compound game is a winning game. However, game B is a winning game instead of a losing game as specified in the proposed game[14]. Actually, the trick employed in game B is relatively simple – no matter whether the starting capital for game B is an odd or even number, the resultant capital is becoming and subsequently maintaining as an odd number with the advancement of number of game played. In order to demonstrate the idea, it begins the game by using an odd number, for instance, *$9*, as starting capital for game B. With the advancement of number of games played, the capital $C(t)$ becomes "*9→15→21→27→33→⋯*", resulting in a winning game. If game B starts with an even-number starting capital, for instance, *$10*, with the advancement of number of games played, the capital $C(t)$ becomes "*10→3→9→15→21→⋯*", which also results in a winning game. Therefore, no matter the starting capital is an odd or even number, game B is always a winning game. There is no doubt that playing game B alone offers higher returns than playing the compound game, which is reflected in Figure 6.

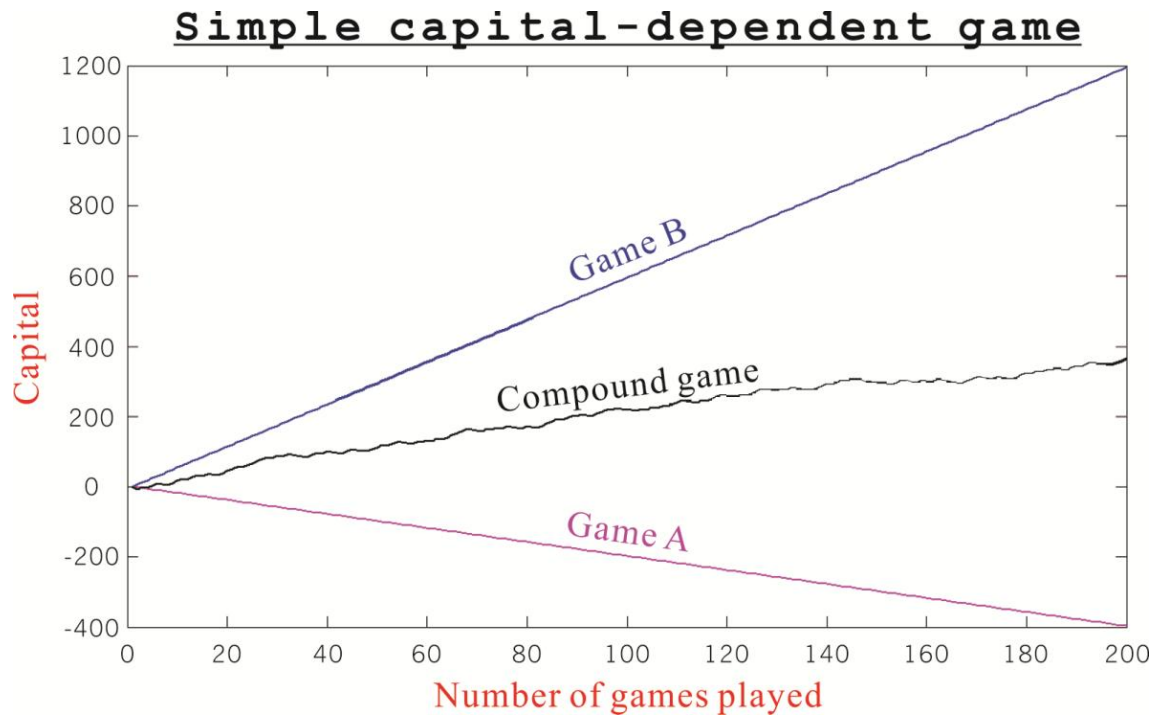

**Figure 6**: *Simple capital-dependent game*

The phenomenon generated by such a simple capital-dependent game is much similar to that of *scheme #3* (Figure 5(b)), which should be treated as a trivial case. In other words, such an effect cannot be treated as the paradoxical effect at all. Therefore, the paradoxical effect cannot be simply created by replacing the original game with a primitive version. Unfortunately, the proposed simple capital-dependent game[14] is failed to reproduce the analogous paradoxical effect. The Parrondo's paradox is caused by manipulating the probability distribution of individual losing games to form a winning compound game[17,18].

These eight different combinations of two winning and/or losing games can be included into the same probability space, as shown in Figure 7.

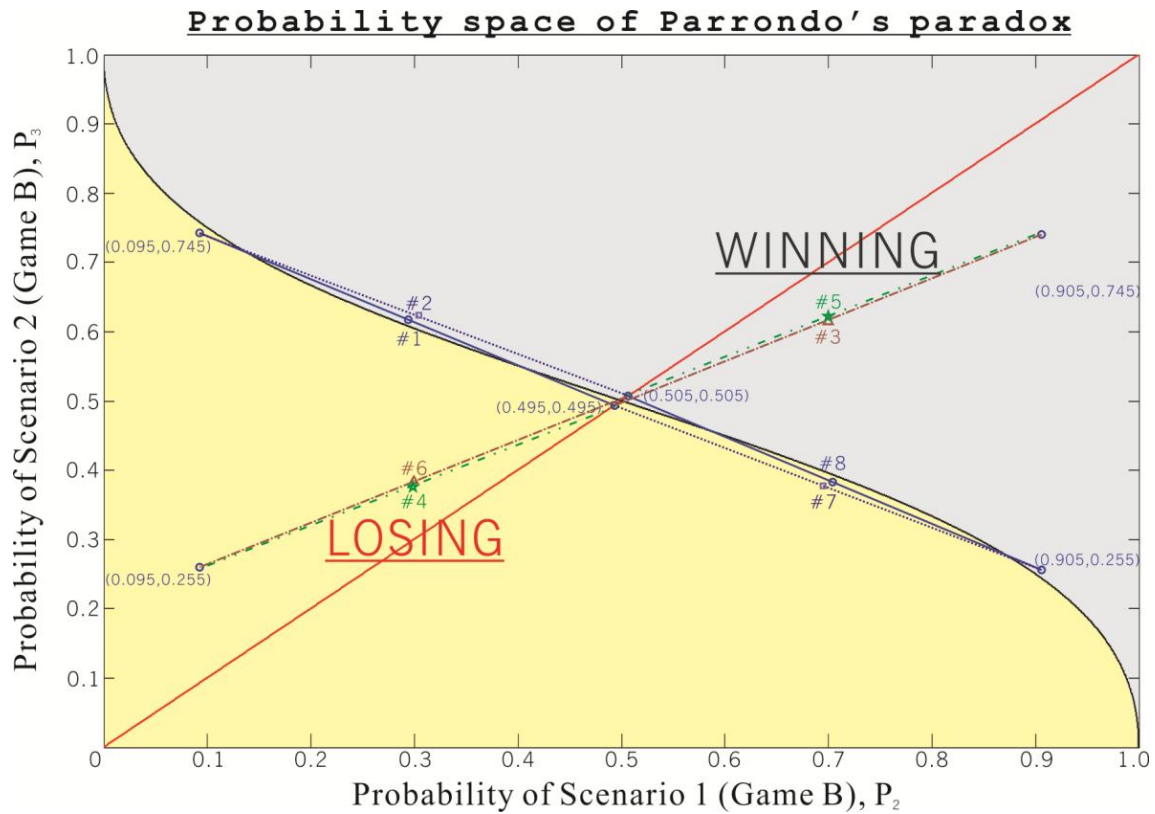

**Figure 7**: *Probability space containing all probabilities of eight combinations*

Due to the feature of point symmetry, the analysis can be simply restricted to one side of the probability space, that is, schemes *#1*, *#2*, *#4* and *#6*, in which *scheme #1* is the original capital-dependent Parrondo's paradox.

In *scheme #2*, game A is a winning game as its winning probability is greater than $\frac{1}{2}$, and game B is a losing game, which is exactly the same as that of *scheme #1*. In this case, the compound game is also a winning game as the center point of the connecting line between these two points is falling inside the *winning region*. In *scheme #4*, game A is a losing game, which is the same as that of *scheme #1*, and game B is also a losing game as the probabilities of both scenarios are falling inside the *losing region*. There is no doubt that the compound game is also a losing game. Finally, in *scheme #6*, game A is a winning game as that of *scheme #2*, and game B is a strong losing game as that of *scheme #4*. The resultant compound game is also a losing game.

Based on the one-sided analysis, it is able to determine the results on the other side. The compound games of *scheme #3* (reversed *#6*), *scheme #5* (reversed *#4*), *scheme #7* (reversed *#2*), *scheme #8* (reversed *#1*) are winning, winning, losing and losing games, respectively. After conducting a series of simulations (Figure 5), the results for various schemes are summarized in Table 2. In summary, schemes *#1* and *#8* are able to produce very strong paradoxical effect. Schemes *#2* and *#7* are capable of producing relatively strong paradoxical effect. For the remaining schemes, *#3* to *#6*, are failed to generate any paradoxical effects. These schemes can be regarded as trivial cases, labeled as "N/A".

**Table 2:** *Summary of results corresponding to different schemes*

| Scheme | Description | Paradoxical effect |
|---|---|---|
| #1 | Lose + Lose = Win | Very strong |
| #2 | Win + Lose = Win | Strong |
| #3 | Lose + Win = Win | N/A |
| #4 | Lose + Lose = Lose | N/A |
| #5 | Win + Win = Win | N/A |
| #6 | Win + Lose = Lose | N/A |
| #7 | Lose + Win = Lose | Strong |
| #8 | Win + Win = Lose | Very strong |

## Methods

*Modified probability curve*

It is able to observe the fact that the Parrondo's paradox can be reproduced as long as there exists a connecting line of two selected points of probabilities across the curve boundary with two points located in the *losing region* and the middle section of the connecting line falling in the *winning region*. Therefore, it is possible to modify the probability curve based on this observation. The simplest modification is done by changing the value of predefined integer number $M = 5$. After the modification, the resultant function becomes equation (7).

$$p_2 = \frac{(1-p_3)^4}{(1-p_3)^4 + p_3^4}. \tag{7}$$

The modified probability curve (solid black line) and its original probability curve (dash grey line) are shown in Figure 8.

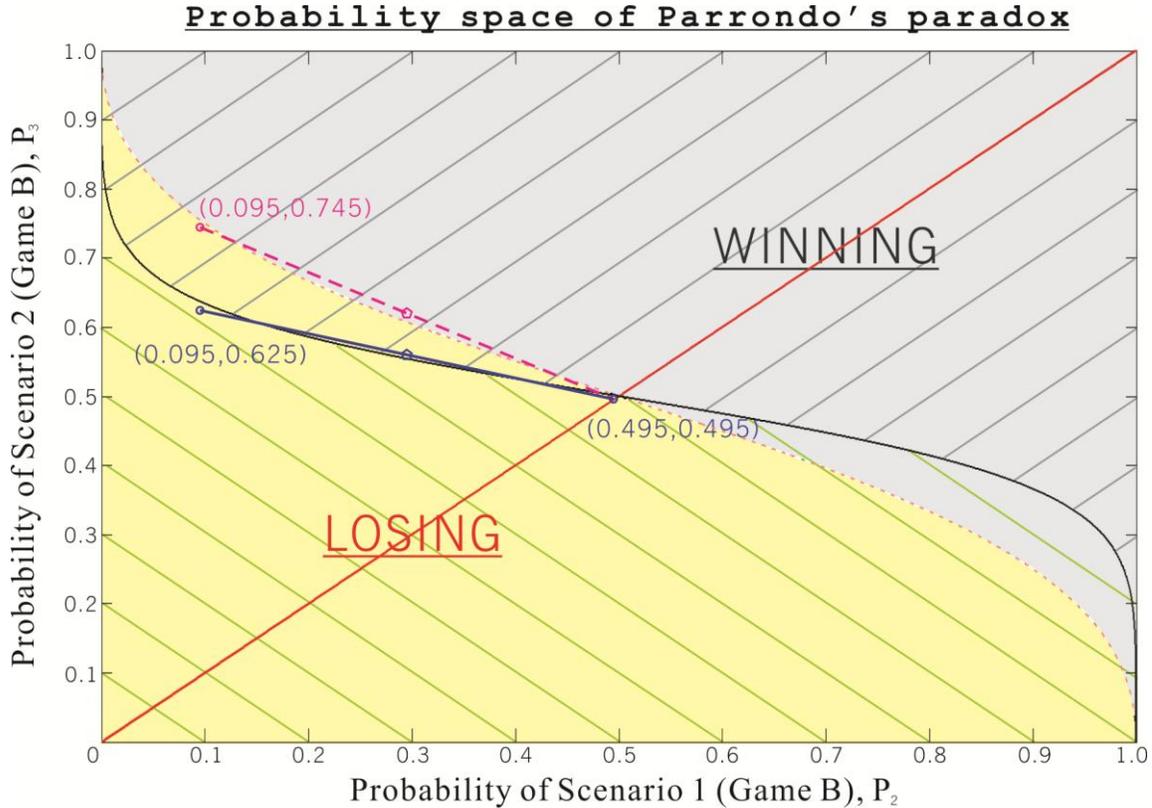

**Figure 8**: *Probability space of original and modified Parrondo's paradox*

Similarly, the curve divides the entire probability space into two regions, the *winning region* (yellow-line shaded area) and *losing region* (grey-line shaded area). The original and modified probability curves have one property in common, that is, both of them are symmetric about the intersection point of curve boundary and straight line that represents game A. Due to this specific property, the probability of game A remains unchanged, that is, $p_1 = 0.495$. On the opposite, the original probabilities of game B, ($p_2$, $p_3$) = (*0.095*, *0.745*), is no longer feasible as it is falling inside the *winning region*. By adjusting the point to the new location, ($p_2$, $p_3$) = (*0.095*, *0.625*), the paradoxical effect can be reproduced for this modified case. As shown in Figure 8, the probability of game A is falling in the *losing region* as usual; the probability of game B is also falling inside the *losing region*. However, there is a certain region of connecting line located in the *winning region*. By manually controlling the location of compound game, *i.e.*, modify the value of mixing parameter, the resultant compound game can also be a winning game. Based on these data, and keeping the remaining parameters unchanged, the simulation of this case can be produced as shown in Figure 9.

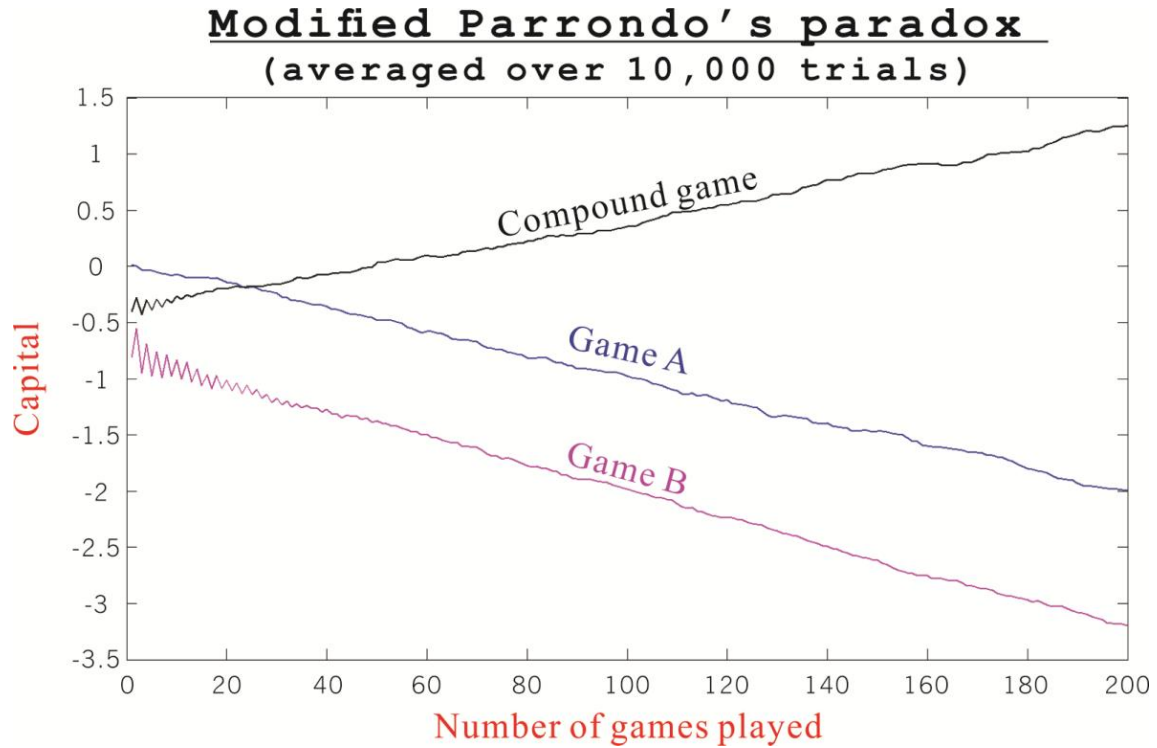

**Figure 9**: *Modified capital-dependent Parrondo's paradox*

*Non-linear combination of two games*

In previous cases, the compound game is formed in terms of a convex linear combination of two individual games. In reality, the combination of these two individual games, game A and game B, can also be non-linear. In the following case, the concept of non-linear combination of two individual games is demonstrated to be outperformed the original linear combination of two games. The previous case, whereas predefined integer $M = 5$, is used in this demonstration. All these three non-linear combinations together with the linear combination are shown in Figure 10.

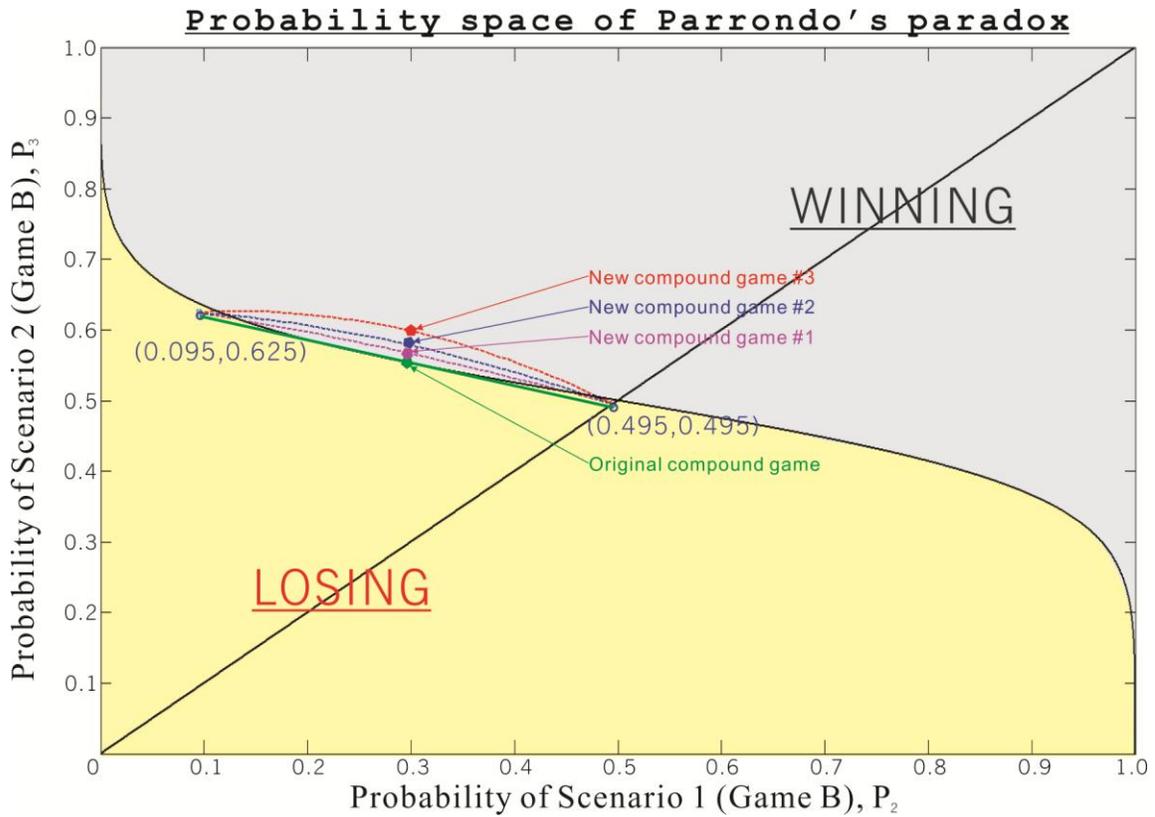

**Figure 10**: *Linear/non-linear combinations in probability space*

The original linear combination of two games is represented by a green solid line. By introducing one mixing parameter $\gamma$, it is possible to control the probability of the compound game along the line (also mentioned in previous section). The non-linear combinations of these two games are expressed in terms of dash lines in Figure 10. The functions of these lines are determined based upon these two probability points. Finally, only the middle points of these functions are selected as the probabilities of the compound game.

The simulation is produced based upon the same parameters as previous case. Only in this case, deciding which game to be played is no longer depending on the mixing parameter $\gamma$. Instead, two probabilities of compound game, $p_{c1}$ and $p_{c2}$, are firstly determined. As both compound game and game B of the capital-dependent Parrondo's paradox are condition-based game, it is possible to directly employ the probabilities of compound game under the paradigm of game B. As shown in Figure 11, the similar paradoxical effects can be reproduced even the compound game is formed in a way of non-linear combination. Furthermore, one additional intriguing feature can be observed from the simulation result, that is, the capital is proportional to the distance between the selected probabilities of compound game and the curve boundary.

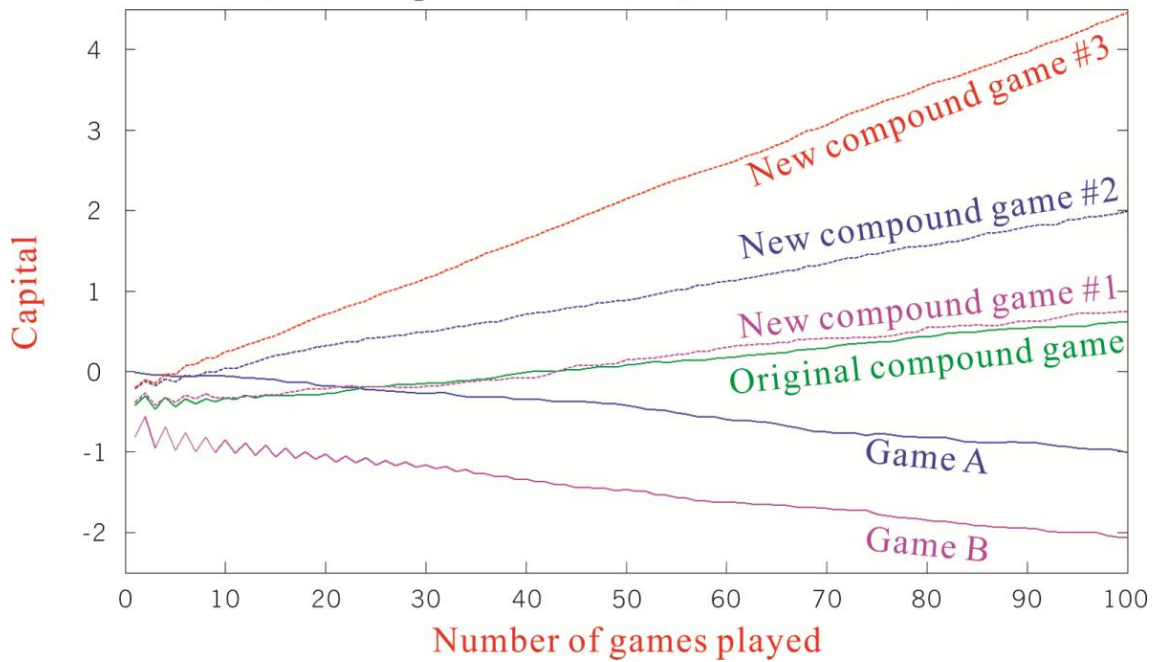

**Figure 11**: *Simulation of non-linear combinations*

*Concluding remarks*

The paper investigates whether the combinations of two winning and/or losing games are capable of generating possible paradoxical effects. It is shown that the identical paradoxical effect cannot be simply reproduced by employing a relative simple capital-dependent game. In reality, the phenomenon generated by the Parrondo's paradox, can be explained by placing all probabilities in a probability space. The paradoxical effect can be produced by either modifying the probability boundary or arranging two winning and/or losing games in a way of linear/non-linear combination.

## Author contributions

J.J.S. conceived and designed the study and Q.W.W. performed the simulations. All authors reviewed the manuscript.

## Additional information

**Competing financial interests**: The authors declare no competing financial interests.